\documentclass[english,aps,superscriptaddress,showkeys,showpacs,prepri,twocolumn]{revtex4}
\usepackage[T1]{fontenc}
\pdfoutput=1
\usepackage[letterpaper]{geometry}
\usepackage{units}
\usepackage{bbding}
\usepackage{amsmath}
\usepackage{amssymb}
\usepackage{graphicx}
\usepackage{esint}
\usepackage[colorinlistoftodos,prependcaption,textsize=tiny]{todonotes}

\usepackage[utf8]{inputenc}
\usepackage{lmodern} 

\makeatletter

\providecommand{\tabularnewline}{\\}

\@ifundefined{textcolor}{}
{%
 \definecolor{BLACK}{gray}{0}
 \definecolor{WHITE}{gray}{1}
 \definecolor{RED}{rgb}{1,0,0}
 \definecolor{GREEN}{rgb}{0,1,0}
 \definecolor{BLUE}{rgb}{0,0,1}
 \definecolor{CYAN}{cmyk}{1,0,0,0}
 \definecolor{MAGENTA}{cmyk}{0,1,0,0}
 \definecolor{YELLOW}{cmyk}{0,0,1,0}
}

\usepackage{doi}
\usepackage{hyperref}

\makeatother

\usepackage{babel}
\begin{document}

\title{Effect of Scrape-Off-Layer Current on Reconstructed Tokamak 
Equilibrium}

\author{J. R. King}
\affiliation{Tech-X Corporation, 5621 Arapahoe Ave. Boulder, CO 80303, USA}

\author{S. E. Kruger}
\affiliation{Tech-X Corporation, 5621 Arapahoe Ave. Boulder, CO 80303, USA}

\author{R. J. Groebner}
\affiliation{General Atomics, PO Box 85608, San Diego, CA 92186–5608, USA}

\author{J. D. Hanson}
\affiliation{Auburn University, 206 Allison Labs, Auburn, AL 36849, USA}

\author{J. D. Hebert}
\affiliation{Auburn University, 206 Allison Labs, Auburn, AL 36849, USA}

\author{E. D. Held}
\affiliation{Department of Physics, Utah State University, 4415 Old Main Hill, Logan, UT 84322, USA}

\author{J. R. Jepson}
\affiliation{Department of Physics, Utah State University, 4415 Old Main Hill, Logan, UT 84322, USA}


\date{January 3rd, 2017}
\begin{abstract}
Methods are described that extend fields from reconstructed equilibria to include
scrape-off-layer current through extrapolated parametrized and experimental
fits. The extrapolation includes both the effects of the toroidal-field and
pressure gradients which produce scrape-off-layer current after recomputation
of the Grad-Shafranov solution. To quantify the degree that inclusion of
scrape-off-layer current modifies the equilibrium, the
$\chi$-squared goodness-of-fit parameter is calculated for cases with and
without scrape-off-layer current. The change in $\chi$-squared is found to be
minor when scrape-off-layer current is included however flux surfaces are 
shifted by up to $3\;cm$. The impact on edge modes of
these scrape-off-layer modifications is also found to be small and the
importance of these methods to nonlinear computation is discussed.
Published version: Phys. Plasmas 24, 012504 (2017) [\url{http://dx.doi.org/10.1063/1.4972822}]
\end{abstract}



\maketitle



  \newcommand{\vect}[1]{ \mathbf{#1}}
  \newcommand{\defn}{ \equiv}

  \newcommand{\lp}{\left(}
  \newcommand{\rp}{\right)}
  \newcommand{\lb}{\left[}
  \newcommand{\rb}{\right]}
  \newcommand{\la}{\left<}
  \newcommand{\ra}{\right>}

  \newcommand{\vf}{ \vect{f}}

  \newcommand{\vx}{\vect{x}}
  \newcommand{\vq}{\vect{q}}
  \newcommand{\vB}{\vect{B}}
  \newcommand{\vJ}{\vect{J}}
  \newcommand{\vA}{\vect{A}}
  \newcommand{\vE}{\vect{E}}
  \newcommand{\vV}{\vect{V}}
  \newcommand{\vF}{ \vect{F} }	
  \newcommand{\vU}{ \vect{U} }	
  \newcommand{\ddp}{\grad \cdot \Pi}
  \newcommand{\specheat}{\gamma_h}

  \newcommand{\grad}{\vect{\nabla}}
  \newcommand{\curl}[1]{\grad \times #1 }
  \newcommand{\dive}[1]{\grad \cdot #1 }
  \newcommand{\vdg}{\left(\vV \cdot \grad \right)}
  \newcommand{\bdg}{\left(\vB \cdot \grad \right)}
  \newcommand{\divV}{\grad \cdot \vV_1}
  \newcommand{\divVp}{\left( \grad \cdot \vV \right)}

  \newcommand{\dt}[1]{\frac{\partial #1}{\partial t}}
  \newcommand{\Dt}[1]{\frac{d #1}{dt}}
  \newcommand{\dpsi}[1]{\frac{\partial #1}{\partial \psi}}
  \newcommand{\dpsisq}[1]{\frac{\partial^2 #1}{\partial \psi^2}}
  
  \newcommand{\jac}{{\mathcal{J}}}
  \newcommand{\jaci}{{\mathcal{J}}^{-1}}
  \newcommand{\Pp}{ P^\prime }				
  \newcommand{\Vp}{V^\prime}
  \newcommand{\Vpp}{V^{\prime\prime}}
  \newcommand{\Vpo}{ \frac{V^\prime}{4 \pi^2}}
  \newcommand{\norm}{ P^\prime }		
  \newcommand{\RR}{ \psi }			
  \newcommand{\vR}{ \grad \RR }		
  \newcommand{\C}{ C }				
  \newcommand{\vC}{ \vect{\C} }		
  \newcommand{\vK}{ \vect{K} }			
  \newcommand{\vRsq}{ \mid \grad \RR \mid^2 }
  \newcommand{\vCsq}{ \C^2 }
  \newcommand{\vKsq}{ K^2 }
  \newcommand{\vBsq}{ B^2 }
  \newcommand{\vrr}{\frac{ \vR}{\vRsq} }
  \newcommand{\vbb}{\frac{ \vB}{B^2} }
  \newcommand{\vcc}{\frac{ \vC}{\vCsq} }
  \newcommand{\vjj}{\frac{ \vJ}{J^2} }
  \newcommand{\vkk}{\frac{ \vK}{\vKsq} }

  \newcommand{\R}{ \psi }
  \newcommand{\T}{ \Theta }
  \newcommand{\Z}{ \zeta }
  \newcommand{\A}{ \alpha }
  \newcommand{\U}{ u }
  \newcommand{\ve}{ \vect{e} }
  \newcommand{\vur}{ \vect{e}^\rho }
  \newcommand{\vut}{ \vect{e}^\Theta }
  \newcommand{\vuz}{ \vect{e}^\zeta }
  \newcommand{\vlr}{ \vect{e}_\rho }
  \newcommand{\vlt}{ \vect{e}_\Theta }
  \newcommand{\vlz}{ \vect{e}_\zeta }
  \newcommand{\gr}{ \grad \R }
  \newcommand{\gt}{ \grad \Theta }
  \newcommand{\gz}{ \grad \zeta }
  \newcommand{\ga}{ \grad \alpha }
  \newcommand{\gu}{ \grad \U }
  \newcommand{\dr}[1]{ \frac{\partial #1}{\partial \R} }
  \newcommand{\dT}[1]{\frac{\partial #1}{\partial \Theta}}
  \newcommand{\dz}[1]{\frac{\partial #1}{\partial \zeta}}
  \newcommand{\dU}[1]{\frac{\partial #1}{\partial \U}}
  \newcommand{\drs}[1]{ \frac{\partial^2 #1}{\partial \R^2} }
  \newcommand{\dTs}[1]{\frac{\partial^2 #1}{\partial \Theta^2}}
  \newcommand{\drt}[1]{\frac{\partial^2 #1}{\partial \R \partial \Theta}}
  \newcommand{\dzs}[1]{\frac{\partial^2 #1}{\partial \zeta^2}}
  \newcommand{\grr}{ g^{\R \R} }
  \newcommand{\grt}{ g^{\R \Theta} }
  \newcommand{\grz}{ g^{\R \zeta} }
  \newcommand{\gtz}{ g^{\Theta \zeta} }
  \newcommand{\gtt}{ g^{\Theta \Theta} }
  \newcommand{\gzz}{ g^{\zeta \zeta} } 
  \newcommand{\ri}{ \frac{1}{R^2} }
  \newcommand{\fr}{ \lp \R \rp}
  \newcommand{\frt}{ \lp \R, \T \rp}
  \newcommand{\frtz}{ \lp \R,\T,\Z \rp}

  \newcommand{\fluxav}[1]{\la #1 \ra}
  \newcommand{\thetaav}[1]{\la #1 \ra_\T}

\newcommand{\cramplist}{
        \setlength{\itemsep}{0in}
        \setlength{\partopsep}{0in}
        \setlength{\topsep}{0in}}
\newcommand{\cramp}{\setlength{\parskip}{.5\parskip}}
\newcommand{\zapspace}{\topsep=0pt\partopsep=0pt\itemsep=0pt\parskip=0pt}

\section{Introduction}
\label{sec:introduction}

Understanding of tokamak plasmas has greatly benefited from the
separation of time scales inherent in strongly magnetized plasmas.  
All orderings of strongly-magnetized plasma equations show that the magnetic
pressure and line-bending contributions from $\vJ \times \vB$ and the
force-density from $\nabla p$ are the largest terms in the center-of-mass
momentum equation.
Assuming symmetry and nested flux surfaces allows one to derive the
Grad-Shafranov equation~\cite{shafranov71} that describes the lowest-order
steady-state fields. This solution is conventionally described as the plasma
equilibrium.  Perturbations that drive the plasma away from equilibrium launch
Alfv\'en waves that quickly act to restore the plasma to
equilibrium~\cite{Jenkins:2010bz}.  Transport occurs on slower time scales,
while symmetry-breaking macroscopic instabilities occur on intermediate time
scales between the transport and the stiff Alfv\'en time scale.

This paradigm well-describes the experimental phenomenology, and solutions to
the Grad-Shafranov equation are routinely calculated hundreds of times per
plasma discharge to provide a measure of the location of magnetic-flux
surfaces.  As such, these solutions are indispensable to controlling the plasma
and interpreting diagnostic data.  Thus Grad-Shafranov theory may be considered
as one of the most successful applications of tokamak-plasma theory.
Measurement of both the macro- and micro-scopic perturbations confirms the
small-fluctuation assumption embedded in the theory: tearing modes, for
example, have perturbed magnetic energies four orders of magnitude below the
equilibrium stored magnetic energy~\cite{sauter}, and ion-temperature-gradient-driven turbulence
produces density fluctuations less than one percent of the equilibrium
density~\cite{white2008measurements}. The small-fluctuation hierarchy coupled
with slow temporal evolution permits the common approach of extended-MHD
modeling about a Grad-Shafranov equilibrium (e.g. Ref.~\cite{King16qh}), as
opposed to modeling the symmetric fields with the full extended-MHD
equations (e.g.  Ref.~\cite{Ferraro09}). The effects encompassed by
extended-MHD are variable: extended-MHD refers to models beyond resistive MHD
that include some combination of anisotropic thermal conduction and stresses
\cite{sovinec04}, two-fluid evolution \cite{Sovinec10}, finite-Larmor-radius
closures \cite{Schnack06,King11}, and/or advanced drift-kinetic-equation closures \cite{held15}.

%

One challenge to simulating small fluctuations about an equilibrium state is
that errors in the equilibrium can be on the order of the perturbation
magnitude. This issue is recognized by Grimm et al.~\cite{grimm76} with the
first `mapping code', a code that employs numerical maps to transfer fields
from the spatial discretization of an equilibrium code onto the spatial
discretization of another code, where the second code typically assesses
stability or the evolution of nonlinear perturbations. Despite considerable
effort to increase the accuracy of mapping codes, mapping errors are difficult,
if not impossible, to completely eliminate.  In practice, extended-MHD codes,
such as \textsc{nimrod}~\cite{sovinec04,Sovinec10} and
\textsc{m3d-c1}~\cite{jardin2007high,Ferraro10}, recompute the Grad-Shafranov
equilibrium with their native spatial discretizations to circumvent these
errors. Burke et al.~\cite{Burke10} discusses an example of the impact of
mapping errors.  In that work, the \textsc{nimrod} extended-MHD code is
benchmarked with the linear-MHD codes \textsc{gato}~\cite{bernard1981gato} and
\textsc{elite}~\cite{Snyder02} on peeling-ballooning modes in tokamak
equilibria. \textsc{gato}, like other global linear-MHD codes, uses mapped
equilibria while \textsc{elite} uses Miller equilibria~\cite{miller98} that
effectively act to provide the same accuracy as re-solving the Grad-Shafranov
equation. In \textsc{nimrod} calculations with mapped equilibria, better
agreement with \textsc{gato} is obtained, whereas when the \textsc{nimeq}
code~\cite{Howell14} is used to recompute the Grad-Shafranov equation for the
\textsc{nimrod} initial condition, better agreement with \textsc{elite} is
obtained.  

\begin{figure}
\begin{center}
\includegraphics[width=0.5\textwidth]{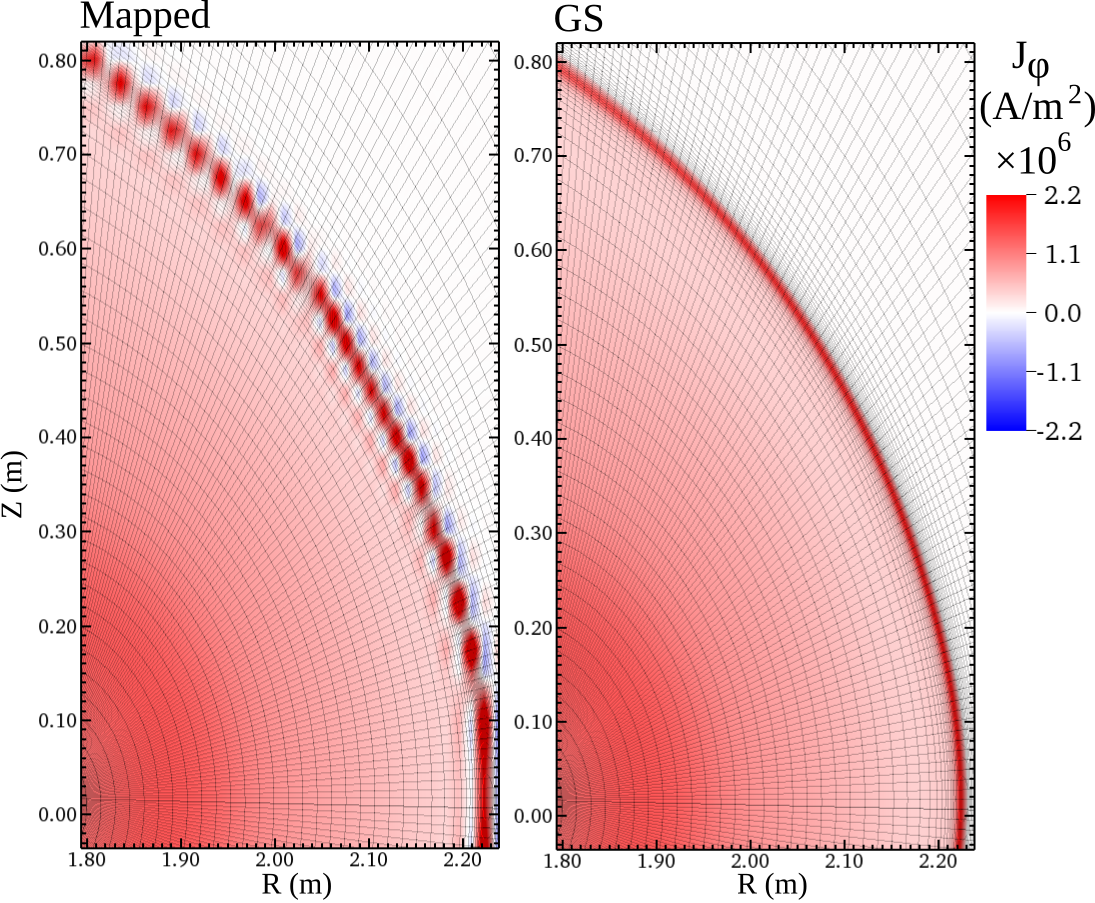}
\caption{ [color online]
Toroidal current on the upper half of the outboard midplance from equilibria
that are mapped (left) and that use a recomputed Grad-Shafranov solution
(right). From a reconstruction of DIII-D shot 145098 at 1800 ms.
}
\label{fig:jphiHR}
\end{center}
\end{figure}

While this example demonstrates the importance of high-accuracy Grad-Shafranov
solutions that are recomputed within a code's native spatial discretization, a
further example is provided with Fig.~\ref{fig:jphiHR}. The figure shows a
comparison of the toroidal current density from both a mapped equilibrium 
along with a case where the Grad-Shafranov solution
is recomputed with \textsc{nimeq}. This high-resolution computation is
performed with a $72\times512$ finite-element grid with bi-quartic elements.
The bumps in the mapped solution roughly correspond to the resolution of the
$129\times129$ grid used during the reconstruction. Computations of edge-mode
growth rates with the mapped equilibria give markedly different results
relative to using the solution recomputed with \textsc{nimeq}.

\begin{figure}
\begin{center}
\includegraphics[width=0.35\textwidth]{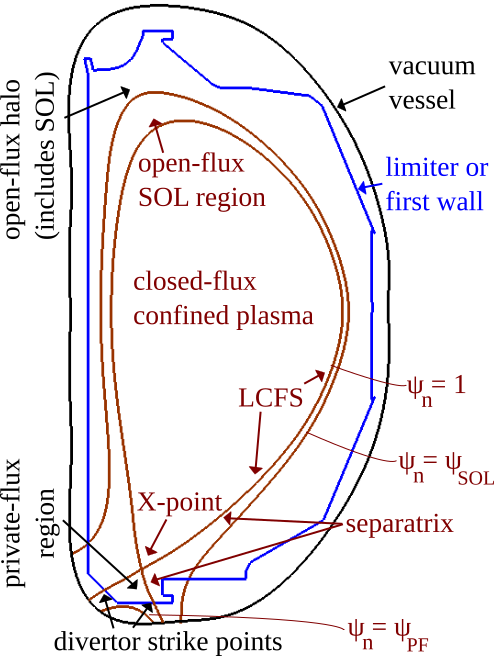}
\caption{ [color online]
Diagram of topological regions of a diverted, lower-single-null shot within the
DIII-D tokamak.
}
\label{fig:contours}
\end{center}
\end{figure}

The work of Burke et al.\ considered the simple case of an artificial
peeling-ballooning equilibrium with exclusively closed and nested-flux surfaces
without diverted magnetic topology. A diagram of the topology associated with
cases considered in this
work, a diverted lower-single-null shot within the DIII-D tokamak, is shown in
Fig.~\ref{fig:contours} along with labels of the terminology used for the
topological regions and contours. Operation with a divertor implies the
presence of a separatrix field line that separates the open- and closed-flux
regions. The surface immediately inside the separatrix is the last-closed-flux
surface (LCFS). The separatrix and LCFS are indistinguishable from a practical
standpoint, aside from the separatrix field lines that extend to the divertor
strike point. The open field lines between the LCFS and the first wall define
the halo region~\cite{strauss2004mhd,Kruger:2005dc}, and in the modeling of
Burke et al.  this region was approximated as a closed-field-line region.
Traditionally in linear-MHD modeling, this halo region is treated as a vacuum;
i.e., it has neither current nor plasma density. In extended-MHD modeling, the
term halo region is used instead to denote that it is a cold-plasma region,
capable of containing current, surrounding the hot, closed-field-line region. 


The goal of this work is to relax the vacuum constraint of the halo region
while remaining consistent with the measurements that are used to produce the
reconstruction.  
For model fidelity to experiments, it is important to carefully consider
the vacuum assumption in the halo region relative to the dynamics of study.
Edge modeling is likely sensitive to current and flows present in the
scrape-off-layer (SOL) region~\cite{Groth:2007ik}, a subregion of the halo that
directs the energy and particle exhaust from the hot plasma onto the divertor.
Most published simulations to date are initialized from
equilibrium that do not have current or flow in the halo region.  This is largely
because this current is typically not included in reconstructed equilibria
from the highly successful, workhorse \textsc{efit}
code~\cite{Lao:1985ww,Lao:1985wa}.  Without SOL current, reconstructions must
have one of two undesirable properties: 1) either there is an artificial
constraint on the current which must smoothly vanish at the separatrix or 2)
there is a current discontinuity at the separatrix.  The former constraint
leads to the incorrect edge profiles, whereas the latter impacts convergence --
particularly for high-order spatial discretizations such as those employed by
\textsc{nimrod} and \textsc{m3d-c1}.  Importantly, cases with either constraint
fail to include the measured profiles outside the LCFS.

In this paper, we
discuss our method for adding SOL current to equilibrium
reconstructions generated by \textsc{efit}. 
One notable case that uses this method is the DIII-D QH-mode modeling in
Ref.~\cite{King16qh}.  The difficulties with the current and flow discontinuities and a
desire for more accurate modeling of the QH-mode instabilities motivated the
development of our methods that include SOL profile gradients.
This paper begins by reviewing details of
the \textsc{efit} reconstructions, and discussing important metrics for
determining the quality of the reconstructions in Sec.~\ref{sec:efitreview}.
The methods for recomputing equilibria with separatrices along with the
details on how we add SOL current to these new equilibria  are described in
Sec.~\ref{sec:algorithms}. Example cases that use these methods are introduced
in Sec.~\ref{sec:examples}. We then compare the new and original equilibria
using synthetic diagnostics that are similar to those used to constrain the
original reconstructed equilibrium in Sec.~\ref{sec:diagnostics}.  Finally, we
examine the minor modification to the linear growth rates of peeling-ballooning
modes (PBMs) by the addition of SOL current in Sec.~\ref{sec:pbm} before
making concluding remarks in Sec.~\ref{sec:conclusions}.
\section{Overview of \textsc{efit} equilibrium reconstructions}
\label{sec:efitreview}

Determination of the experimental configuration of tokamak plasmas
has become essential for understanding and optimizing stability and confinement
in fusion research devices.  Reconstruction of the experimental equilibrium
from a combined set of magnetic, temperature and density measurements is
computed by minimizing the error between modeled and observed signals.  This
technique, now routine, was pioneered with the \textsc{efit} code which
originally used only magnetic diagnostics external to the first wall as a
constraint~\cite{Lao:1985ww,Lao:1985wa}.  Later, the motional Stark effect
diagnostic~\cite{wroblewski1990motional,rice99} greatly improved the accuracy of the
reconstructions by providing internal measurements of the magnetic and electric
fields.  More recently, measurements of the density and temperature profiles
via Thompson scattering~\cite{carlstrom92,eldon12} and charge exchange recombination
(CER)~\cite{groebner90,chrystal15} spectroscopy provide further
constraints on the pressure profiles~\cite{Lao90}.

To quantify a reconstruction's accuracy, results from \textsc{efit} Grad-Shafranov
solves are applied to a $\chi$-squared test against the experimental measurements.
In other words, the goal is to find a Grad-Shafranov solution that minimizes
\begin{equation}
   \chi^2 = \sum_i{\lp \frac{M_i - C_i}{\sigma_i} \rp^2}
   \label{chi-squared}
\end{equation}
where $M_i$ is the measured signal, $C_i$ is the computed signal, and
$\sigma_i$ is the measurement uncertainty.  In this paper, we perform this
$\chi$-squared calculation to quantify the change that arises from the
numerical errors associated with mapping, our recomputation of the Grad-Shafranov
solution, and the addition of the SOL pressure profiles and associated current
during the recomputation.

Typically, \textsc{efit} reconstructions do not include SOL current and treat
the region between the separatrix and wall, i.e. the halo region, as current
free. This implies that if the plasma discharge has a finite current on the
separatrix, which is often the case for H-mode discharges that contain large
bootstrap and Pfirsch-Schl\"uter currents, there will be a discontinuity at the
LCFS. Equivalently within the context of the Grad-Shafranov equation, the
\textsc{efit} profiles are constrained such that the pressure and toroidal
magnetic field are constant in the halo region.  With finite gradients at the
LCFS, this leads to discontinuities in the first derivatives of these profiles.
These inconsistencies lead to subtle but significant issues when evolving
tokamak-edge unstable cases initialized from \textsc{efit} reconstructions.
For example, in codes that retain the magneto-sonic wave physics both the SOL
current and profile gradients must be included such that the equilibrium is
consistent with force balance. Furthermore during nonlinear computations, any
inconsistencies to the Grad-Shafranov equation, including discretization errors
from discontinuous fields, can launch spurious magneto-sonic waves.

%

%

It must be noted that \textsc{efit} has the ability to include force-free,
poloidal current in the SOL through finite gradients in the toroidal magnetic
flux~\cite{Lao:1991vk,Strait:1991uo}. However, this capability is rarely exercised
and there is no published work on the inclusion of finite pressure gradients
in the SOL.

%
\section{Algorithms: Bounding Contours and SOL Fits}
\label{sec:algorithms}

Initializing a \textsc{nimrod} computation from a reconstruction is a
two-step process that involves mapping from the reconstructed equilibrium, and
then recomputation of the equilibrium.  The mapping
code \textsc{fluxgrid} creates both a partially flux-aligned finite-element mesh
and maps the reconstructed-equilibrium fields onto this mesh.  This
mapped solution is refined through solution of the Grad-Shafranov
equation with the \textsc{nimeq} code~\cite{Howell14}. The solution can
then be iteratively passed between \textsc{fluxgrid} and \textsc{nimeq}
for further grid refinement and Grad-Shafranov solves, as needed. The
Grad-Shafranov solve is formulated as a boundary-value problem where the
boundary condition is specified by the value of $\psi$ from the mapped
reconstruction.  This boundary condition constrains both the fields from
the external coils and those generated by the internal plasma that are
also determined by the pressure and toroidal-flux profiles.  With this
method, the recomputed fields closely resemble the reconstructed
versions; however, mapping errors are largely eliminated and the
Grad-Shafranov equation is satisfied up to an input tolerance. 

Relative to the methods described in Ref.~\cite{Howell14}, we employ extensions
that identify the open- and closed-flux regions of the domain in order to apply
the appropriate fitted form of the pressure and toroidal flux profiles. Within
the closed-flux regions, these profiles are specified by \textsc{efit}
reconstructions as a function of normalized flux
($\psi_n = (\psi - \psi_o)/(\psi_x - \psi_o)$ which is zero at the O-point
where $\psi=\psi_o$ and unity on the separatrix where $\psi=\psi_x$).  With the
exception of private-flux regions, regions with $\psi_n<1$ contain closed field
lines and regions with $\psi_n>1$ contain open-field lines. 
Applying these profiles to the domain requires identification
of the closed-flux region. In practice, we find the bounding LCFS contour and
use a simple algorithm that counts the number of crossings of a line that extends from a
finite-element node to the boundary to determine if each node is enclosed by
the contour.  The separatrix contour and location of the extrema of $\psi$ in the
core are allowed to change and must be recomputed after each iteration of the
Grad-Shafranov solve.  As the X-point on the separatrix contour associated with
diverted magnetic topology consists of a stagnation point for field-line
tracing, we instead elect to find a contour vanishingly close to the separatrix
which numerically approximates the LCFS. For
our purposes, we only need to bound the finite-element nodes that are within
the separatrix.  We employ two methods that use field-line tracing of the polodial
field to find this approximate LCFS contour.  The first method uses bisection
of the domain to determine where the field lines transition from closed to open
within a specified tolerance. For the parallel implementation, it becomes an
N-section method where each core is assigned a seed point between the known
closed- and open-field-line locations.  The second method is to use the
\textsc{oculus} code to find the saddle point in $\psi$ that is associated with
the X-point, and a field-line seeded with a vanishingly small offset towards
the O-point is used as the LCFS in order to avoid stagnation of the field-line
tracing near the X-point.  The latter method has the advantage of being
somewhat more robust for high-resolution cases, but the disadvantage of not
being amenable to a straight-forward parallel implementation.

As discussed earlier, \textsc{efit} solutions have zero profile gradients
outside the LCFS.  A cubic-spline fit of this data
is used to evaluate these fields within the LCFS.  Equilibrium generated using
these spline fits with constant profile values outside the LCFS closely match
the solution given by \textsc{efit} while largely eliminating mapping errors.
One goal of this work is to contrast this recomputed solution with a solution
that contains SOL-profile gradients.


SOL-profile gradients and associated currents are included by defining bounding
contours and normalizing the flux as an extension to the methods used to
determine the open- and closed-flux regions.  The SOL region is defined as the
region between the LCFS contour and contours with $\psi_n=\psi_{sol}$ and
$\psi_n=\psi_{pf}$ where $\psi_{sol}$ ($>1$) defines a contour(s) at the edge
of the SOL region and $\psi_{pf}$ ($<1$) defines a contour(s) in the
private-flux region(s) (see Fig.~\ref{fig:contours}).  
The algorithm that
determines these contours is able to handle an arbitrary number of
intersections of this region and the computational boundary of the domain and
thus diverted topology such as double null configurations is tractable.  The
algorithm works as follows: The initial SOL bounding region is assumed to be
the computational boundary.  The algorithm checks the value of $\psi_n$ at each
finite-element node location outside the LCFS but contained within this
`working' SOL region.  If the condition $\psi_{pf}<\psi_n<\psi_{sol}$ is not
satisfied, the location between the wall and the O-point where
$\psi_n=\psi_{sol}$ or $\psi_n=\psi_{pf}$ is identified, and a new contour is
traced. Absent integration error, the new contour is terminated at the
boundary. The section of the domain that does not encircle the O-point is
removed from the `working' SOL region until only nodes with the property
$\psi_{pf}<\psi_n<\psi_{sol}$ remain. This new SOL contour, in addition to the
LCFS contour, bound the SOL region. 

The appropriate pressure and toroidal magnetic flux profiles are defined within
the SOL region via two methods: either by fits from the experimental data (if
available) or through modified bump function fits. Two fits are performed
for each field: one fit for the SOL region immediately outside the separatrix with
$\psi_n>\psi_x$, and one for the private-flux SOL region with $\psi_n<\psi_x$.
The modified bump function uses the form
\begin{equation}
f\left(\psi_n\right)=f_0 Exp\left[\frac{-\Delta^2}{\psi^2_{n,sol} - \psi^2_n} \right]+f_c\;.
\label{bumpftn}
\end{equation}
This function has vanishing derivatives of all orders at its endpoint and thus
ensures that the current goes smoothly to zero at the transition contour
between the SOL region and the current-free region.  With this form, five free
parameters are available ($\psi_{sol}$, $\psi_{pf}$, $f_0$, $f_c$ and
$\Delta$). The values of $\psi_{sol}$ and $\psi_{pf}$ are inputs
that influence the width of the SOL and may be inferred from
experimental measurements. The other
three parameters may be determined either by requiring $C^2$ continuity at the
LCFS or by setting the functional value at $\psi_{sol}$ and enforcing $C^1$
continuity at the LCFS.  The former constraint has the advantage of producing a
$C^1$ smooth current profile, whereas the latter method allows specification of
the density and/or temperature in the current-free regions.  If the fit
requires that the function first reverse the sign of its derivative, a truncated Gaussian
is fit to half of the domain followed by the bump function as is used in
Ref.~\cite{Groebner_JRT}.

\section{Example cases}
\label{sec:examples}

\begin{figure}
\begin{center}
\includegraphics[width=0.5\textwidth]{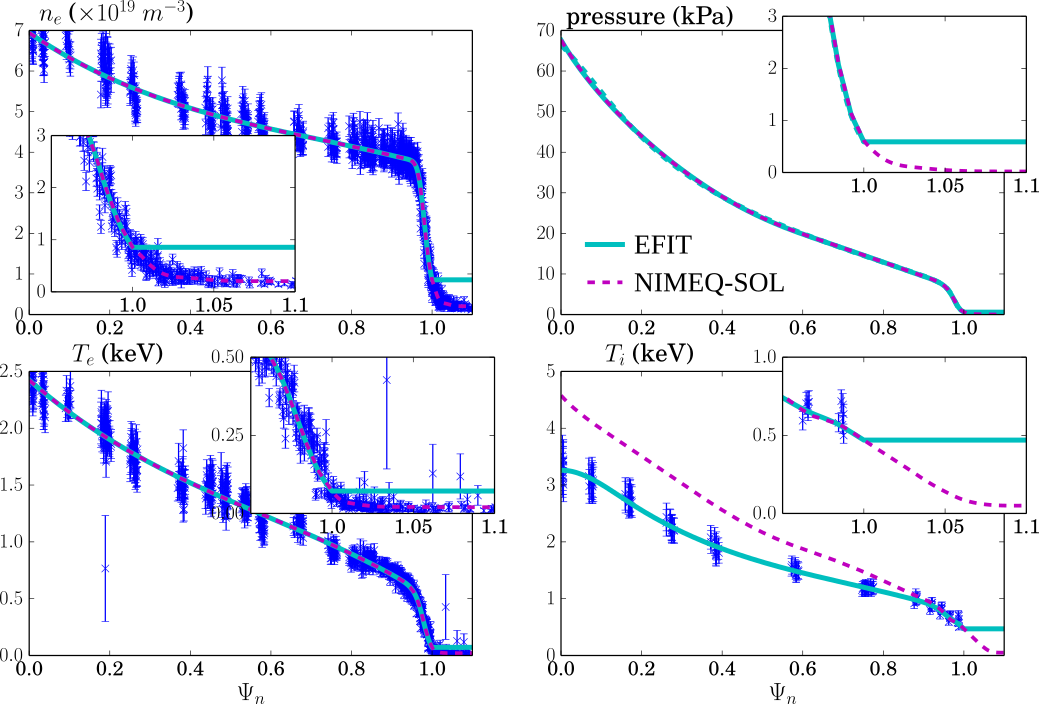}
\caption{ [color online]
Fitted profiles (lines) from the Thompson (electron density and temperature)
and CER (ion temperature) data from shot 160414 at 3025 ms. The solid lines are
fits to raw data inside the LCFS where total pressure is computed from the
measured species' data and includes contributions from energetic particles.
The dashed lines are the fits that include the SOL region and dashed ion
temperature is computed after constraint by a two temperature model and
quasineutrality.
}
\label{prof160414}
\end{center}
\end{figure}

In order to demonstrate and quantify the impact of our methods that recompute
the Grad-Shafranov solution and add SOL-profile gradients, we choose two
specific cases to study in detail: a low SOL-current case and a high SOL-current case.  
The first of these cases is an \textsc{efit} reconstruction of
DIII-D shot 160414 at 3025 ms with profiles as shown in Fig.~\ref{prof160414}.
Profiles are specified as function of normalized flux. The profiles shown are as
included from the \textsc{efit} reconstruction and for fits with SOL-profile
gradients (\textsc{nimeq}-SOL) along with the experimental Thomson and CER
measurements 
\footnote{CER measurements of $T_i$ outside the LCFS are not
included as the signal from the low density plasma in the region is corrupted
by interference from trapped ions within the LCFS.}.
This shot is from an experiment of lithium pellet injection for edge-localized-mode pacing and
the specific time represents the last 20\% of the inter-ELM period.  The
reconstruction contains a relatively cold plasma at the LCFS, and thus a small
pressure gradient in the fitted profiles within the SOL region. This implies
modest SOL current and modification to the resulting equilibria when the SOL
profiles are included in the Grad-Shafranov solve.  In particular, at the LCFS
$T_e=72$ eV, $T_i=470$ eV and $n_e=8.6\times10^{18}$ $m^{-3}$ and in the
current-free region $T_e=20$ eV, $T_i=50$ eV and $n_e=2.1\times10^{18}$
$m^{-3}$ where the latter values only apply to equilibria with SOL-profile
gradients.

\begin{figure}
\begin{center}
\includegraphics[width=0.5\textwidth]{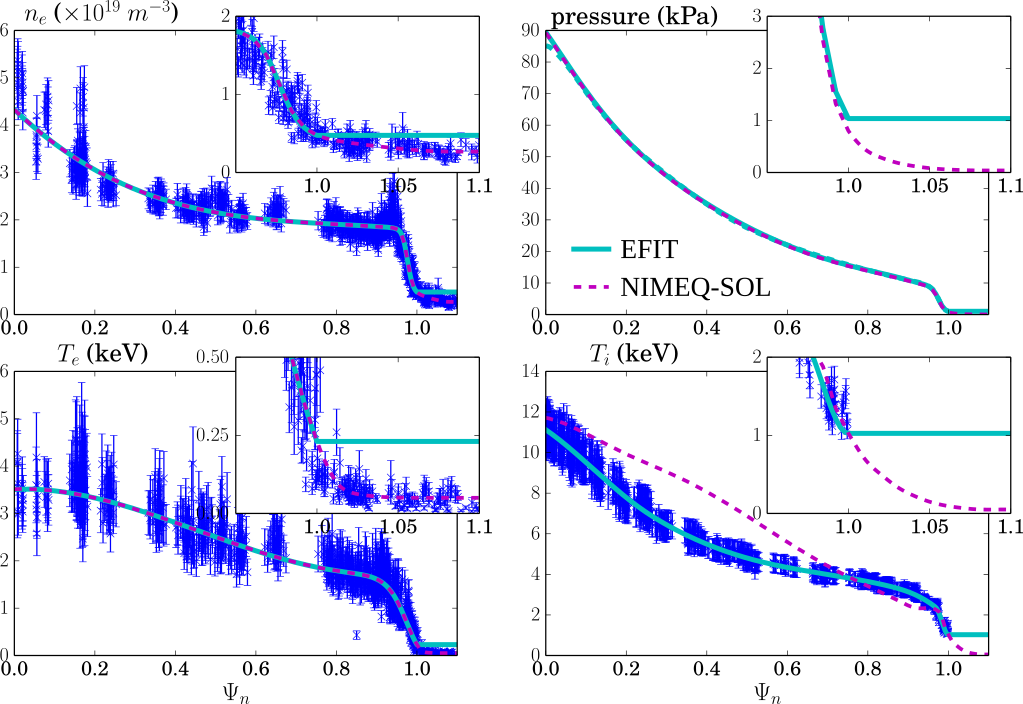}
\caption{ [color online]
Fitted profiles (lines) from the Thompson (electron density and temperature)
and CER (ion temperature) data from shot 145098 at 1800 ms. The solid lines are
fits to raw data inside the LCFS where total pressure is computed from the
measured species' data and includes contributions from energetic particles.
The dashed lines are the fits that include the SOL region and dashed ion
temperature is computed after constraint by a two temperature model and
quasineutrality.
}
\label{prof145098}
\end{center}
\end{figure}

The second case studied is an \textsc{efit} reconstruction of DIII-D shot
145098 at 1800 ms as shown in Fig.~\ref{prof145098}.  Again, the profiles shown
are as included from the \textsc{efit} reconstruction and for fits with SOL-profile 
gradients (\textsc{nimeq}-SOL) along with the experimental Thomson and
CER measurements.  This shot is from a DIII-D QH-mode experiment with ITER-like
shaping during a period with edge harmonic oscillations (low-$n_\phi$
perturbations).  This shot contains a relatively large pressure gradient from
the fitted profiles in the SOL region. Thus relative to the reconstruction from
shot 160414, we expect greater modifications to the equilibria resulting from
the inclusion of the SOL-profile gradients in the Grad-Shafranov solve.  In
particular, at the LCFS $T_e=230$ eV, $T_i=1020$ eV and $n_e=4.8\times10^{18}$
$m^{-3}$ and in the current-free region $T_e=50$ eV, $T_i=50$ eV and
$n_e=2.7\times10^{18}$ $m^{-3}$ where the latter values only apply to
equilibria with SOL-profile gradients.

As is clear from Figs.~\ref{prof160414} and ~\ref{prof145098}, the
\textsc{nimeq}-SOL ion-temperature profiles do not match the CER measured data.
This is a consequence of a two-temperature model which over-constrains the
profiles.  Specifically, with a two-temperature model assuming quasi-neutrality
($n_e = Z_{i} n_i$) only the electron and ion fluids contribute to the
pressure:
\begin{equation}
      p^{2T}=p_e + p_i =  n_e T_e + \frac{n_e}{Z_i} T_i^{2T}\;.
      \label{eq:2flP}
\end{equation}
Thus only three of the four profiles shown in the figures ($n_e$, $p$, $T_e$,
and $T_i$) can be matched exactly.  Initializing computations with the measured
pressure profile is essential as both the Grad-Shafranov solution and ideal
stability are critically dependent on this profile. Secondary to this, the
electron temperature profile determines the resistivity profile and the density
profile sets Alfv\'en speed (among other collisionality parameters). 

Within the context of an extended-MHD simulation, the ion-temperature profile
typically becomes significant only when a two-fluid model is evolved and thus
is often allowed to vary with respect to experimental measurements.  In 
experiment, the pressure has contributions from both impurities and 
non-Maxwellian, or `hot', ion particles from neutral-beam injection:
\begin{multline}
      p=p_e + p_i + p_{imp} + p_{hot} =  n_e T_e + n_i T_i  \\
      + \sum_j^{imp} n_j T_j + \sum_h^{hot} n_h T_h\;.
      \label{eq:fullPressure}
\end{multline}
Comparing Eqns.~\eqref{eq:2flP} and \eqref{eq:fullPressure}, the ion temperature used
to initialize our simulations includes the contributions from the other species: 
\begin{equation}
      T^{2T}_i= T_i \
      + \sum_j^{imp} \frac{n_j Z_i}{n_e} T_j
      + \sum_h^{hot} \frac{n_h Z_i}{n_e} T_h\;.
      \label{eq:T2T}
\end{equation}
As expected from this relation and as shown in the plots, the
\textsc{nimeq}-SOL ion temperature (equivalent to $T^{2T}_i$ in
Eqn.~\eqref{eq:T2T}) over estimates
the measured ion temperature.  For these cases, $Z_i$ is chosen such that the
edge ion-temperature profile approximately matches the measured data within
Figs.~\ref{prof160414} and \ref{prof145098}
($Z_i=1.25$)~\footnote{In two-fluid modeling, $Z$ is a constant as constrained
by the quasi-neutrality condition and cannot vary in space as is often the case
for tokamak transport modeling}.  Thus computations in this work use the
\textsc{nimeq}-SOL fits for $\psi_n<1$ and only the computations with
SOL-profile gradients use the \textsc{nimeq}-SOL fits where $\psi_n>1$.
Modeling that includes separate species for hot particles and/or impurities is
required to eliminate the discrepancy with the measurements in the
ion-temperature profile, and is thus planned for future study.

\begin{figure*}
\begin{center}
\includegraphics[width=1.0\textwidth]{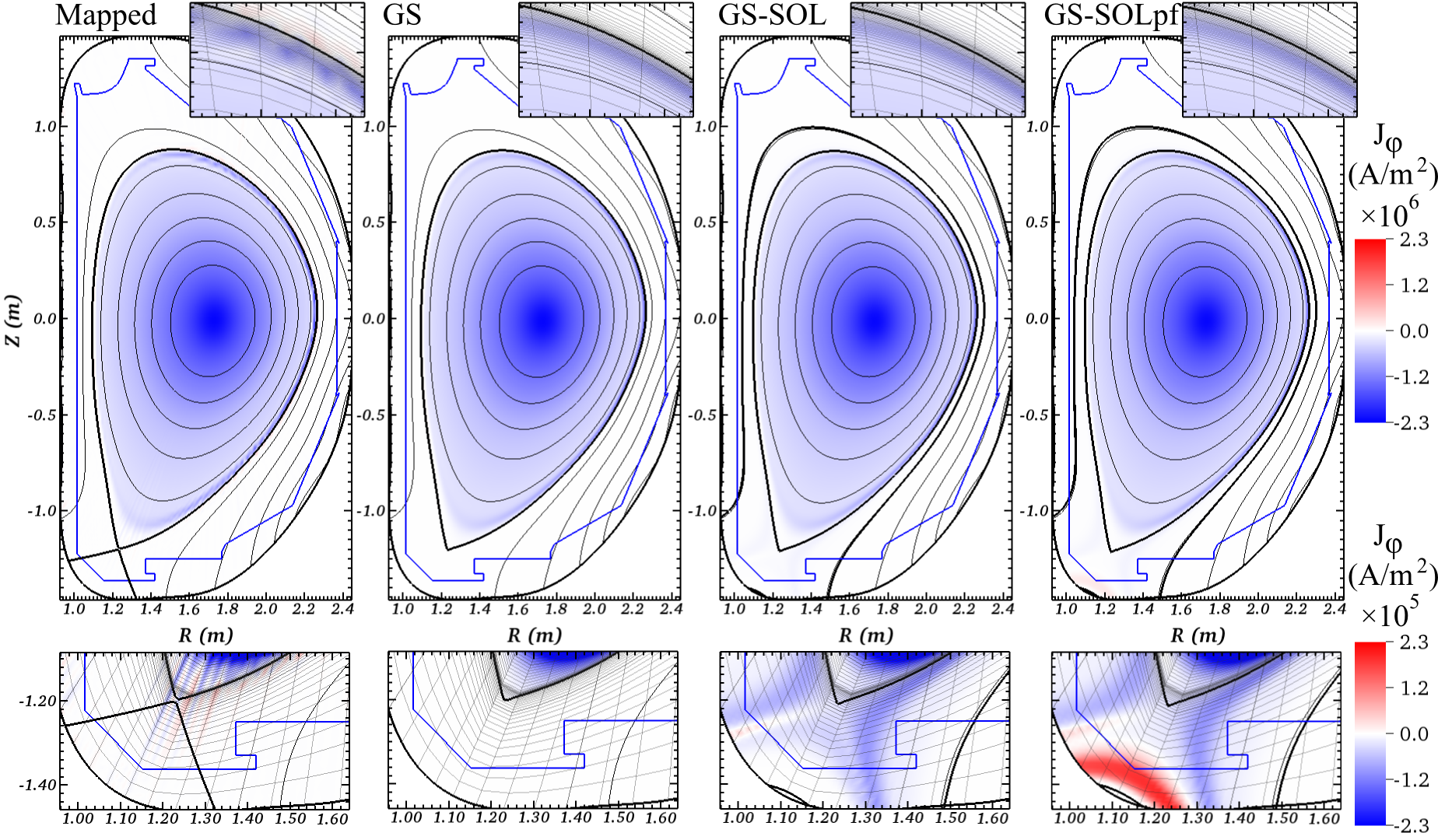}
\caption{ [color online]
Toroidal current density from shot 160414 at 3025 ms plotted with LCFS, SOL (if
applicable) and DIII-D limiter contours for four different cases: a mapped
solution (using a finite-element computation for $\mathbf{B}$ and
$\mathbf{J}$), resolving the Grad-Shafranov equation (GS), and resolving the
Grad-Shafranov equation with two different treatments of the SOL (labeled GS-SOL and
GS-SOLpf) as described in the text. The zoomed plots of the divertor region use
a $10\times$ smaller contour color scale to show current features.
}
\label{fig:jphi160414}
\end{center}
\end{figure*}

\begin{figure*}
\begin{center}
\includegraphics[width=1.0\textwidth]{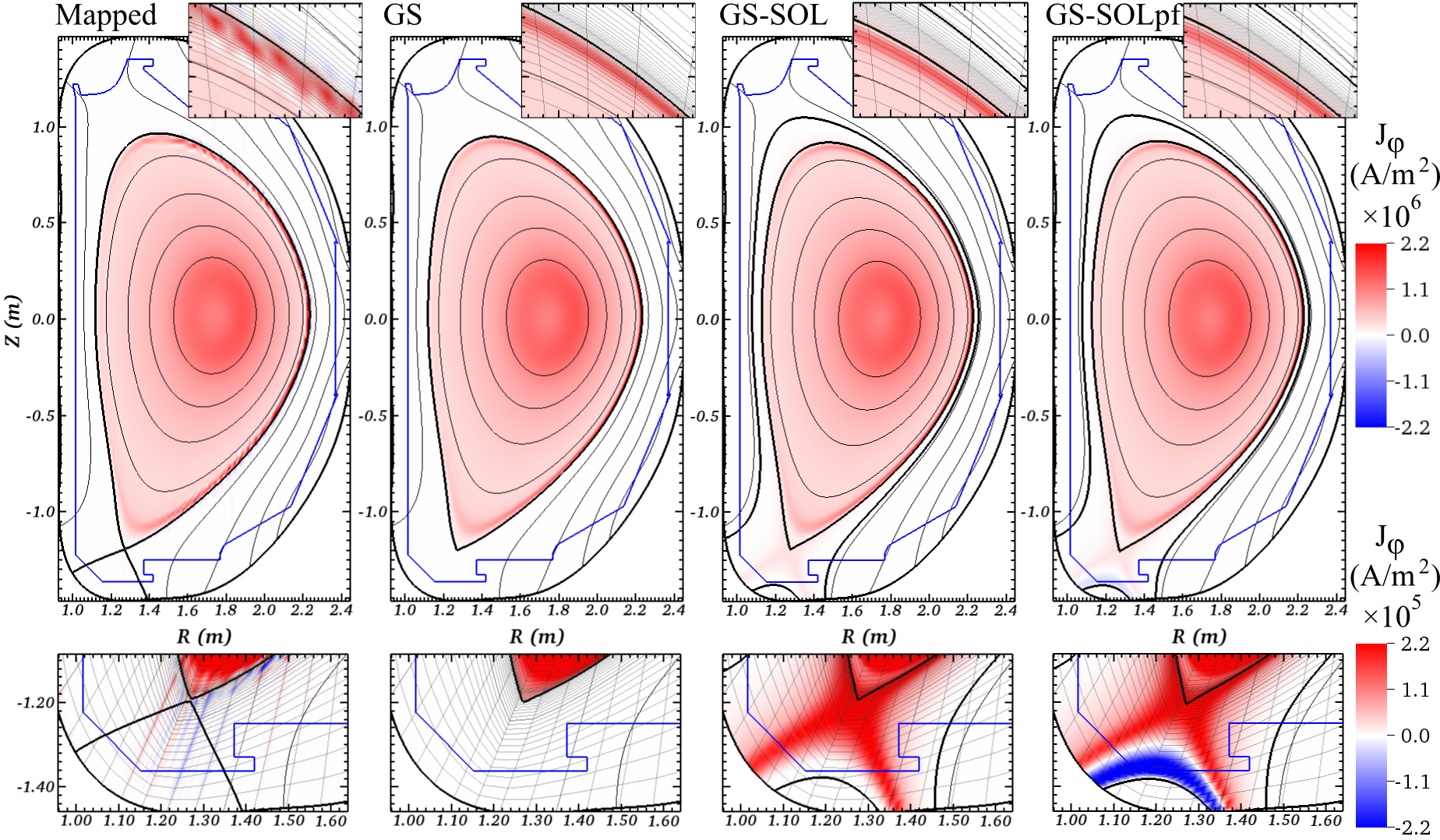}
\caption{ [color online]
Toroidal current density from shot 145098 at 1800 ms plotted with LCFS, SOL (if
applicable) and DIII-D limiter contours for four different cases: a mapped
solution (using a finite-element computation for $\mathbf{B}$ and
$\mathbf{J}$), resolving the Grad-Shafranov equation (GS), and resolving the
Grad-Shafranov equation with two different treatments of the SOL (labeled GS-SOL and
GS-SOLpf) as described in the text. The zoomed plots of the divertor region use
a $10\times$ smaller contour color scale to show current features.
}
\label{fig:jphi145098}
\end{center}
\end{figure*}

To further examine the reduction of mapping errors through the recomputation of
the Grad-Shafranov solution, Figs.~\ref{fig:jphi160414} and
\ref{fig:jphi145098} show the resulting current-density distributions from the
mapped and recomputed (GS) cases without SOL-profile gradients, 
along with two cases with SOL-profile
gradients that are discussed later in this section, based on reconstructions
from shots 160414 and 145098, respectively.  The DIII-D first wall, LCFS and
SOL contours are superimposed into the plotted computational domains. Shot
145098 uses a reversed plasma current relative to 160414, which uses the
standard current orientation for DIII-D. These cases use a $72\times64$
high-order finite-element mesh with bi-quartic elements. For both
reconstructions, the mapped current is clearly distorted and contains numerical
oscillations. This is particularly evident for the edge current shown in the
zoomed insets of the figures and outside the LCFS as shown in the inset figures
that magnify the details near the X-point.  The numerical bumps in the current
profiles inside the LCFS are a result of the mapping and are associated with
the resolution of the \textsc{efit} grid. They are not eliminated but rather
only resolved with enhanced \textsc{nimeq} resolution (see
Fig.~\ref{fig:jphiHR}. While it is possible to
partially circumvent this issue with the closed-flux mapped current by using a
high-resolution \textsc{efit} (see, for example, Ref.~\cite{Aiba07,King16m}),
in practice most \textsc{efit} reconstructions are generated at relatively low
resolution relative to what is required for extended-MHD
computations. The spatial requirements to solve for Grad-Shafranov equilibria
are less stringent than those to solve for 3D MHD perturbations where,
for example, the edge computations presented in Sec.~\ref{sec:pbm} use 
a finite element grid with $72\times512$ with high-order bi-quintic elements. 
The current
outside the LCFS in the mapped case results from the representation of the
discontinuous current profile on \textsc{nimeq}'s $C^0$ finite-element spatial
discretization.  The mapped plots in Figs.~\ref{fig:jphi160414} and
\ref{fig:jphi145098} are generated with finite-element calculations that
compute the poloidal magnetic field and current from the mapped $\psi$ and
$RB_\Phi$ fields.  Alternatively, splines may be used to map the poloidal
magnetic field and current density.  While this method produces smooth, but not
consistent, fields, the derivatives of these fields with \textsc{nimeq}'s $C^0$
finite-element representation are used in extended-MHD calculations. Thus
mapped magnetic fields and current densities with a spline spatial
representation in effect hide, but do not eliminate, the mapping errors.

For the presented cases with SOL current, we use $\psi_{n,sol}=1.1$ and 
$\psi_{n,pf}=0.96$ in the fits to the electron density and temperature profiles from
Thomson scattering measurements and extrapolate the ion temperature to an
assumed $50eV$.  The resulting profiles are shown as the dashed purple line
in Figs.~\ref{prof160414} and \ref{prof145098}.  The half width of the electron
pressure profiles are roughly $3.3\;mm$ and $2.7\;mm$ at the outboard mid-plane
and $7.3\;cm$ and $4.7\;cm$ at the divertor plate for the cases from shots
160414 and 145098, respectively.  This results in SOL widths that are roughly
consistent with the measured half width of the heat-flux during the later half
of the inter-ELM period of DIII-D ELMy H-mode discharges in Ref.~\cite{eich13}.

\begin{figure}
\begin{center}
\includegraphics[width=0.35\textwidth]{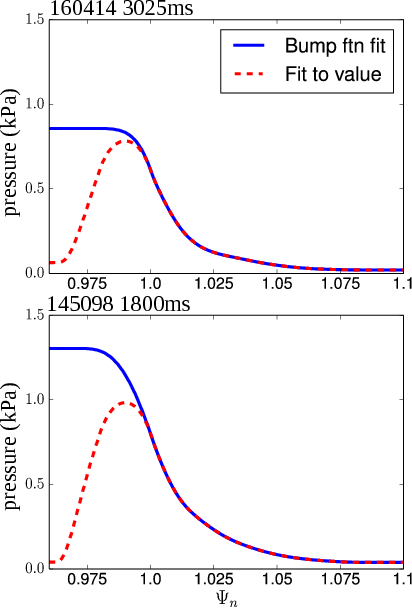}
\caption{ [color online]
Fitted pressure profiles in the SOL and private-flux region with two different
fits, one with bump function fits in the private-flux region (SOL case) and the
second with a fit that specifies a low pressure at the edge of the private-flux
region (SOLpf case), for two different shots, 160414 at 3025ms (top) and 145098
at 1800 ms (bottom).
}
\label{solProfs}
\end{center}
\end{figure}

Currently, we do not have diagnostic information about appropriate profiles for
the private-flux region. Two methods employing the bump function extrapolations
are compared with profiles shown in Fig~\ref{solProfs}. In the first method,
$C^2$ continuity is enforced at $\psi_n=1$ resulting in a high-pressure,
high-density private-flux region. With the second method, $C^1$ continuity is
enforced at $\psi_n=1$ and the values at $\psi_{pf}$ are set to the same as
those at $\psi_{sol}$. In the figures and tables, these cases are referred 
to as GS-SOL and GS-SOLpf, respectively. The profiles generated with the
second method (GS-SOLpf) roughly match the measured heat-flux profile \cite{Groebner_JRT,eich13}. 

In addition to the mapped and recomputed cases without SOL current,
Figs.~\ref{fig:jphi160414} and \ref{fig:jphi145098} plot cases with SOL
currents. With low pressure in the private-flux region the toroidal current
density reverses locally near the divertor strike points after the
Grad-Shafranov solve as shown in the inset figure for the GS-SOLpf case.
Currents near and outside the limiter are likely an artifact of our method.  In
this cold-plasma region, additional terms in the momentum equation become large
(e.g.  interactions with neutrals and the first wall) that are outside the
scope of the Grad-Shafranov equation. As such, the approximation that quatities
are functions of the flux surfaces breaks down. However, at present our focus is
on the effects of the SOL-profile gradients near the LCFS where there is the
potential for interaction with perturbations that originate from inside the
separatrix. With SOL-profile gradients, there is a modest modification (less
than 1\%) to the plasma current. For the cases shown here, we re-normalize the
total current to match the value from the \textsc{efit} reconstruction using
the method of Ref.~\cite{lutjens96}.  Cases with and without current
re-normalization produce similar results, where the $\chi^2$ values discussed
in the Sec.~\ref{sec:diagnostics} are slightly smaller for cases with
re-normalization.

The poloidal current in the SOL that flows into and out of the divertor plate
are solely an effect of the profile gradient in the toroidal magnetic flux
within the SOL. For these cases, a bump-function fit that enforces $C^2$
continuity at $\psi_n=1$ determines the toroidal-magnetic-flux profile in the
SOL. The resulting poloidal currents have a maximum value on the divertor plate
of 4000 and 500 $A/m^2$ for the cases from shots 160414 and 145098,
respectively. This current must be less than the ion saturation current
($J_{max}=n_e e c_s$). With a Deuteron ion species, a conservative calculation of
the ion saturation current (using the vacuum temperatures and densities from
each case) is over $2\times10^4$ $A/m^2$ for both cases.

\section{Synthetic diagnostics}
\label{sec:diagnostics}

Analysis between each nonlinear Grad-Shafranov-solve iteration 
provides a confirmation
that the macroscopic quantities of the equilibrium (e.g. total current,
toroidal flux and internal energy) are invariant between the mapped equilibrium
and the new solution.  In the interest of showing that our methods which recompute
the Grad-Shafranov equilibrium and add SOL profiles and current only minimally
impact the relative agreement with measurements, we use a more advanced
comparison that computes a $\chi^2$ value.  The Python code \textsc{nimnostics} is used
to calculate a $\chi^2$ from the shots and sub-cases previously
discussed. \textsc{nimnostics} models the magnetic coils, MSE and Thomson
scattering through local evaluations of the fields where linear
interpolation is used between finite-element nodes. Thus the boundary of the
domain is chosen as the approximate vacuum vessel instead of the limiter in order to
encompass the magnetic coil locations for comparison. 

\begin{table}
\caption{
Values of $\chi^2 /N$, where $N$ is the number of measurements, for each
diagnostic for the reconstruction, change in plasmas current relative to the
\textsc{efit} value (-1178772 Amps), and change in the X-point and maximum Z value of the
LCFS for shot 160414 at 3025ms.
\label{tab:chisq160414}}
\centering{}%
\begin{tabular}{lcccc}
\hline
$\chi^2 /N$ & mapped & GS & GS-SOL & GS-SOLpf \tabularnewline
\hline
Thom. $T_e$ & 22.3 & 23.4 & 4.80 & 4.15 \tabularnewline
Thom. $n_e$ & 19.4 & 20.5 & 4.07 & 3.33 \tabularnewline
CER $T_i$   & 6.98 & 6.96 & 6.74 & 6.84 \tabularnewline
MSE         & 1.49 & 1.49 & 1.46 & 1.47 \tabularnewline
Mag. Coils  & 0.61 & 0.63 & 0.82 & 0.70 \tabularnewline
\hline
$\Delta$s   & mapped & GS & GS+SOL & GS+SOLpf \tabularnewline
\hline
$\Delta I/I_0$ & 6.95$\times 10^{-5}$ & 7.97$\times 10^{-4}$ & 3.22$\times 10^{-7}$ & 3.22$\times 10^{-7}$ \tabularnewline
$\Delta \mathbf{r}_{xpt}$ (cm)  & N/A & ref. & 0.72 & 1.07 \tabularnewline
$\Delta \mathbf{r}_{zmax}$ (cm) & N/A & ref. & 0.35 & 0.19 \tabularnewline
\hline
\end{tabular}
\end{table}

\begin{table}
\caption{
Values of $\chi^2 /N$, where $N$ is the number of measurements, for each
diagnostic for the reconstruction, change in plasmas current relative to the
\textsc{efit} value (1063788 Amps), and change in the X-point and maximum Z value of the
LCFS for shot 145098 at 1800ms.
\label{tab:chisq145098}}
\centering{}%
\begin{tabular}{lcccc}
\hline
$\chi^2 /N$ & mapped & GS & GS-SOL & GS-SOLpf \tabularnewline
\hline
Thom. $T_e$ & 60.9 & 61.7 & 7.77 & 6.99 \tabularnewline
Thom. $n_e$ & 2.87 & 5.22 & 11.4 & 9.93 \tabularnewline
CER $T_i$   & 10.2 & 10.3 & 19.7 & 19.6 \tabularnewline
MSE         & 1.14 & 1.13 & 1.13 & 1.13 \tabularnewline
Mag. Coils  & 1.65 & 1.60 & 4.57 & 3.27 \tabularnewline
\hline
$\Delta$s   & mapped & GS & GS+SOL & GS+SOLpf \tabularnewline
\hline
$\Delta I/I_0$ & 9.86$\times 10^{-5}$ & 5.46$\times 10^{-3}$ & -3.57$\times 10^{-7}$ & -3.57$\times 10^{-7}$ \tabularnewline
$\Delta \mathbf{r}_{xpt}$ (cm)  & N/A & ref. & 0.86 & 0.55 \tabularnewline
$\Delta \mathbf{r}_{zmax}$ (cm) & N/A & ref. & 2.82 & 2.12 \tabularnewline
\hline
\end{tabular}
\end{table}

Summaries of the $\chi^2$ value by diagnostic and case for the reconstructions from
shots 160414 and 145098 are shown in Tabs.~\ref{tab:chisq160414} and
\ref{tab:chisq145098}, respectively.  In both shots, the mapped and recomputed
equilibria without SOL-profile gradients (GS cases) are roughtly equivalent
indicating that the equilibrium resulting from the recomputation is
substantially similar.  For the cases with modest SOL-profile gradients 
and currents (shot 160414), the $\chi^2$ values for a given
measurement either decreases and the deviations of the LCFS contour are modest
or are roughly the same as the mapped and GS cases.  However, cases from the
shot with large SOL current and profile gradients exhibit mixed results
with some measurements decreasing in $\chi^2$ (Thomson electron temperature) 
and others increasing in $\chi^2$ (Thomson electron density,  MSE and coils) 
when comparing
relative to the mapped and GS cases. The deviation of the LCFS contour is
also as much as 2 cm on the upper side of the contour. This deviation
is apparent in the inset figures of Fig.~\ref{fig:jphi145098}.

\begin{figure*}
\begin{center}
\includegraphics[width=1.0\textwidth]{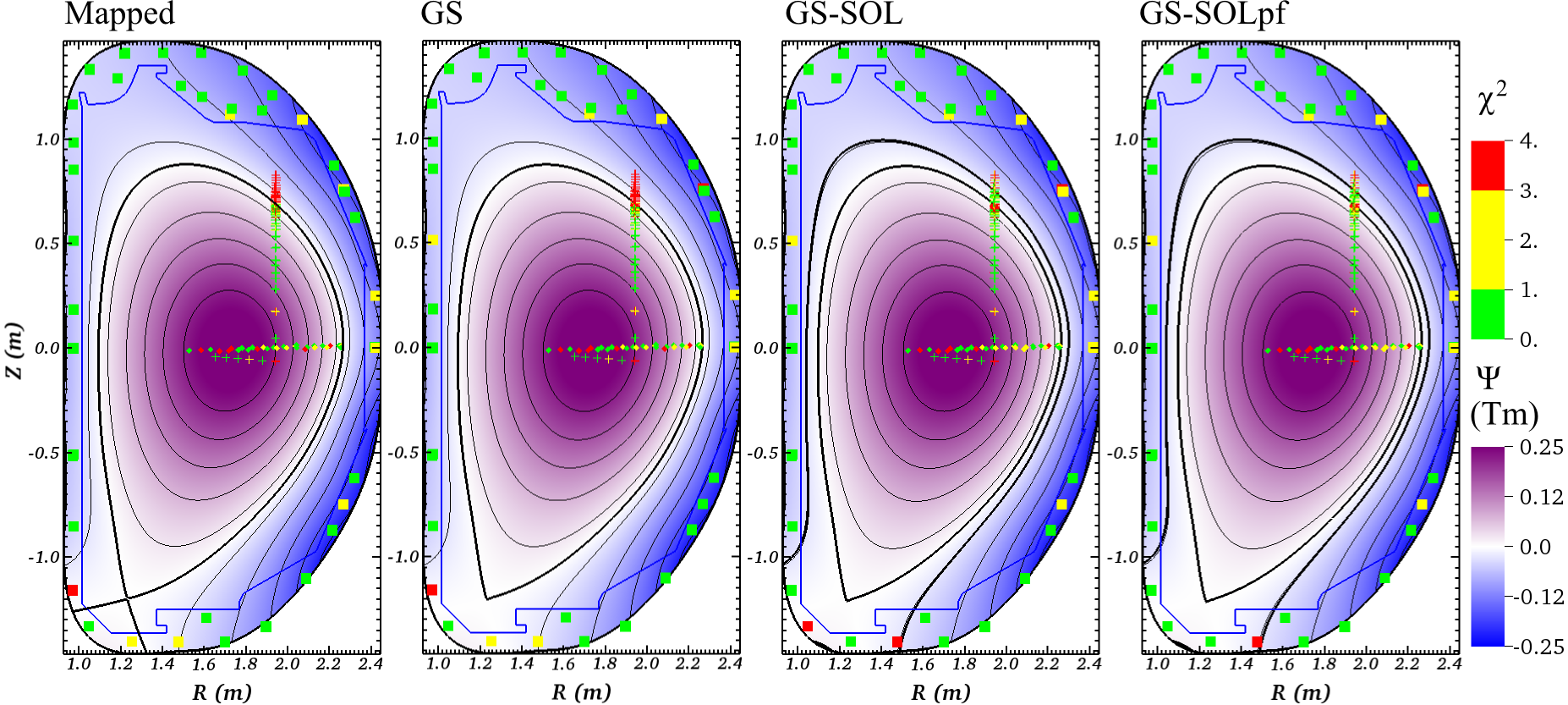}
\caption{ [color online]
Values of $\chi^2$ from different local measurements (crosses are from Thomson
measurements of electron density, diamonds are from MSE measurements, and
squares are from coil measurements of the poloidal magnetic field) plotted with
a color plot of the $\psi$ solution and LCFS, SOL (if applicable) and DIII-D
limiter contours from shot 160414 at 3025 ms for four different cases: a mapped
solution (using a finite-element computation for $\mathbf{B}$ and
$\mathbf{J}$), resolving the Grad-Shafranov equation (GS), and resolving the
Grad-Shafranov equation with two different treatments of the SOL (GS-SOL and
GS-SOLpf) as described in the text.
}
\label{fig:chisq160414}
\end{center}
\end{figure*}

\begin{figure*}
\begin{center}
\includegraphics[width=1.0\textwidth]{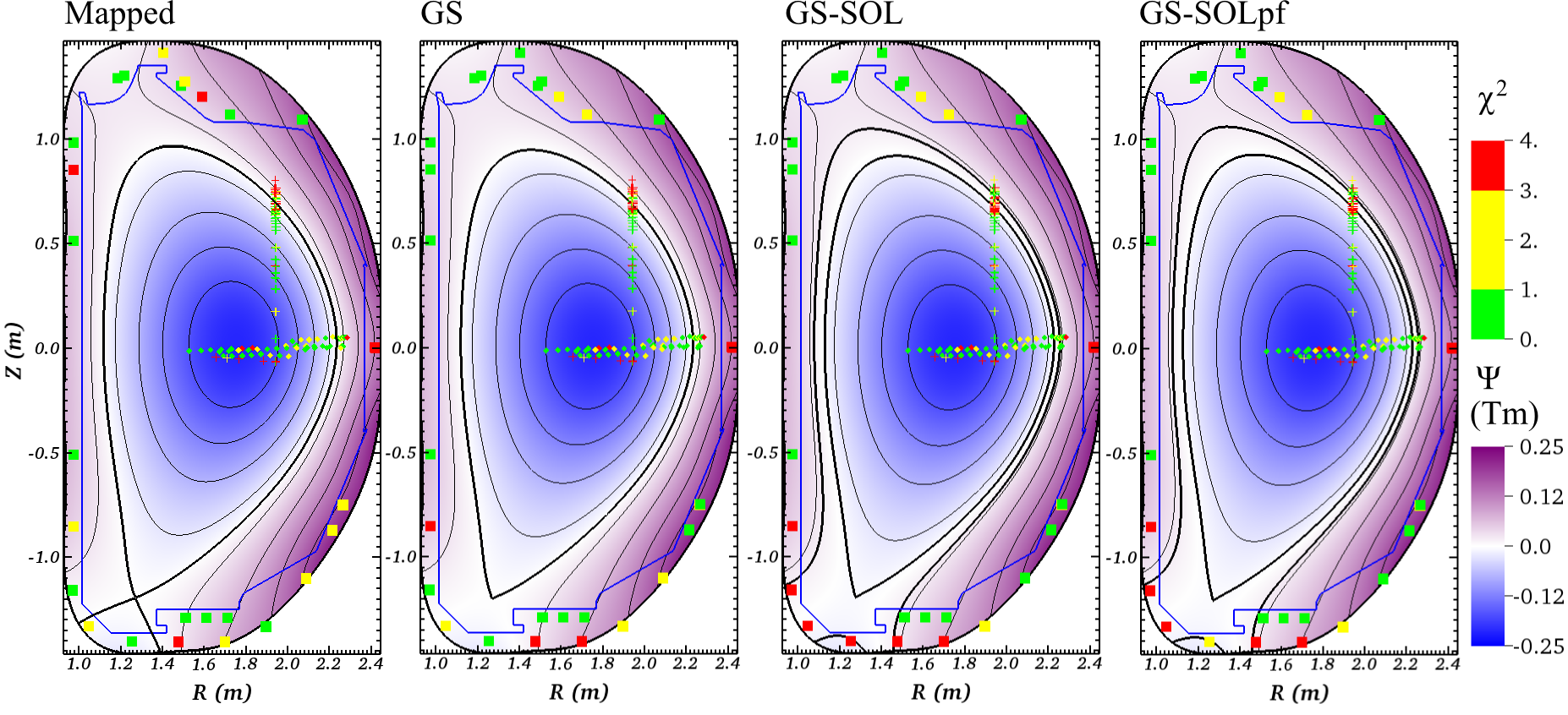}
\caption{ [color online]
Values of $\chi^2$ from different local measurements (crosses are from Thomson
measurements of electron density, diamonds are from MSE measurements, and
squares are from coil measurements of the poloidal magnetic field) plotted with
a color plot of the $\psi$ solution and LCFS, SOL (if applicable) and DIII-D
limiter contours from shot 145098 at 1800 ms for four different cases: a mapped
solution (using a finite-element computation for $\mathbf{B}$ and
$\mathbf{J}$), resolving the Grad-Shafranov equation (GS), and resolving the
Grad-Shafranov equation with two different treatments of the SOL (GS-SOL and
GS-SOLpf) as described in the text.
}
\label{fig:chisq145098}
\end{center}
\end{figure*}

In order to examine the source of the changes in $\chi^2$ in detail,
Figs.~\ref{fig:chisq160414} and \ref{fig:chisq145098} show the local values of
$\chi^2$ from Thomson electron density, MSE and magnetic coils measurements for
our four different cases on shots 160414 and 145098, respectively. With shot
160414 the aggregate $\chi^2$ values are improved or comparable for every
diagnostic when comparing the cases with SOL gradients to those without (as
seen in Tab.~\ref{tab:chisq160414}).  We note that $\chi^2$ values are
particularly improved with SOL-profile gradients for the Thomson measurements,
and Fig.~\ref{fig:chisq160414} shows that this is a result of improved
agreement in the SOL region. The results are mixed for shot 145098 (as seen in
Tab.~\ref{tab:chisq145098}). While the aggregate $\chi^2$ value for the
Thompson electron temperature profile improves when SOL-profile gradients are
included, the $\chi^2$ value for the Thomson electron density profile is
degraded. Examination of Fig.~\ref{fig:chisq145098} shows that while the
$\chi^2$ values in the SOL region are smaller when the SOL-profile gradients
are included, the values near the LCFS become large consistent with the
approximately 2 cm movement of the separatrix line relative to the cases
without SOL-profile gradients. Additionally, the $\chi^2$ values for the
magnetic-coil measurements become marginally larger near the divertor region as
these values are affected by the inclusion of toroidal current in this region.

\begin{figure}
\begin{center}
\includegraphics[width=0.5\textwidth]{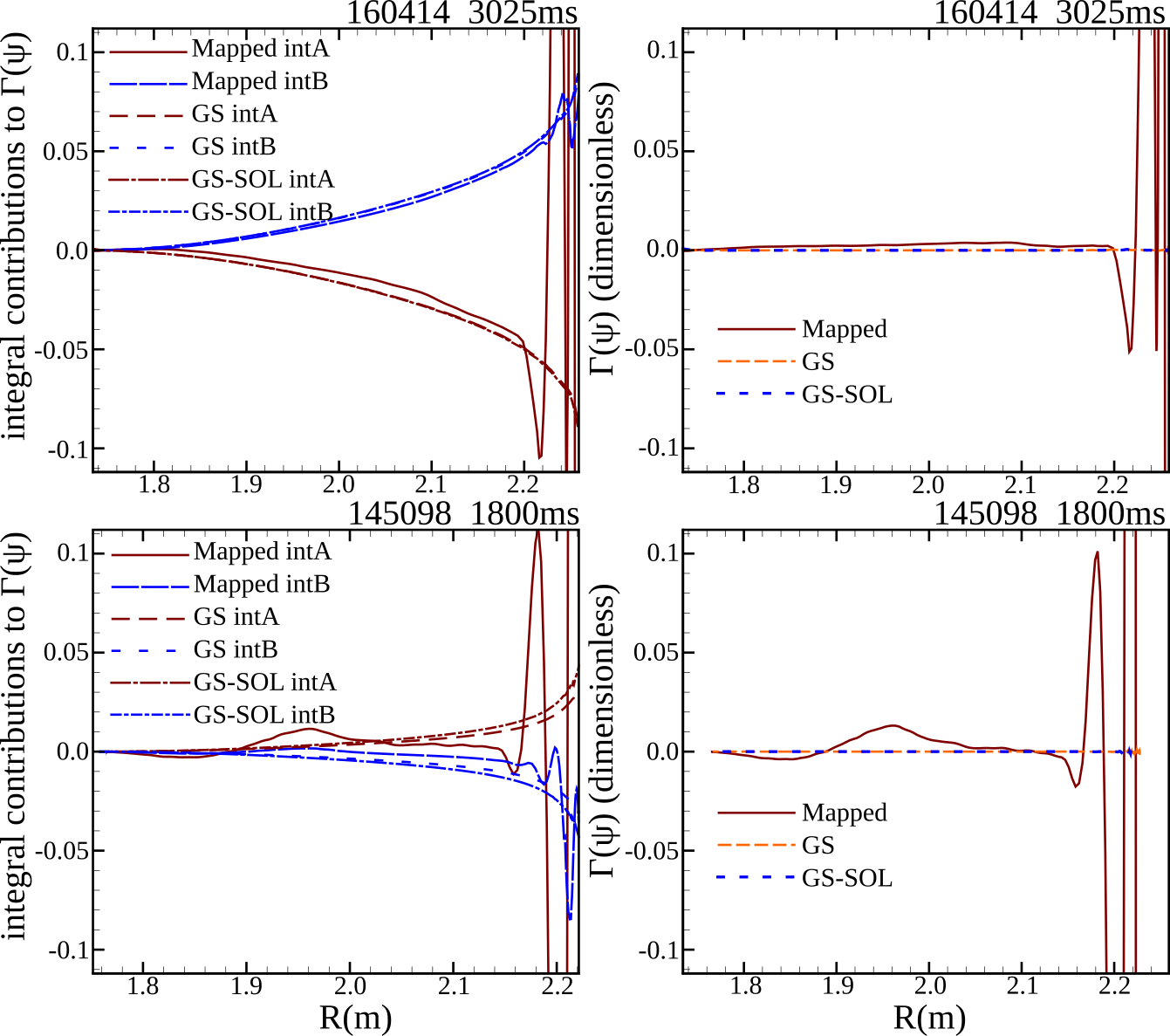}
\caption{ [color online]
The flux-surface-averaged geometrical quantity of Eqn.~\eqref{eq:ramosint} from
Ref.~\cite{ramos15}, $\Gamma (\psi)$ (right), and decomposition by integral
contributions from the two integrals in the expression (left) which are
referred to as intA and intB, respectively. Figures from both example cases, shot
160414 (top) and 145098 (bottom), are shown. The integral vanishes
analytically, and this behavior is only reproduced numerically with
recomputation of the Grad-Shafranov solution.
}
\label{fig:integrals}
\end{center}
\end{figure}

As an additional test of the quality of the equilibrium, we calculate the
flux-surface-averaged geometrical quantity discussed in Ref.~\cite{ramos15},
\begin{multline}
\Gamma (\psi)=2 \frac{d}{d\psi} \oint dl \mathbf{B}\cdot \nabla R^2 \\
+ \oint \frac{dl}{B} [2 \mathbf{b} \cdot \nabla (\nabla \psi \cdot \nabla ln R^2)
                     +\nabla \psi \cdot \nabla (\mathbf{b} \cdot \nabla ln |B|) ]
\label{eq:ramosint}
\end{multline}
where $\mathbf{b}=\mathbf{B}/|B|$. Figure \ref{fig:integrals} shows the result
of this calculation and the decomposition of the expression in
Eqn.~\eqref{eq:ramosint} into separate contributions from the two integrals in
the expression (referred to as intA and intB, respectively) for both shots examined
in our studies.  This integral is known to vanish analytically \cite{ramos15e},
however this behavior is only reproduced numerically with
recomputation of the Grad-Shafranov solution. Consistent with the relatively
large motion of the flux surfaces ($2-3\;cm$) in the large SOL-current case (145098), 
the contributing integrals for this case differ slightly between the GS and
GS-SOL cases where only the latter includes the SOL current. The contributing
integrals lie on top of each other for the low SOL-current case (160414).  The
improved equilibria provided by recomputation of the Grad-Shafranov solution
are critical in NIMROD drift-kinetic computations \cite{held15}.  For example,
an accurate account of $\mathbf{b} \cdot \nabla ln |B|$ in latter integral of
Eqn.~\eqref{eq:ramosint} (intB) is essential when assessing how trapped
particles affect parallel closures for NIMROD's fluid system.
\section{Effect on linear peeling ballooning modes}
\label{sec:pbm}

\begin{figure}
\begin{center}
\includegraphics[width=0.4\textwidth]{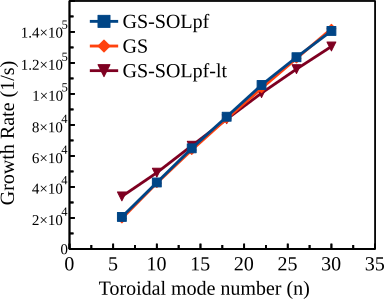}
\caption{ [color online]
Growth rates vs.~toroidal mode from shot 145098 at 1800 ms with and without SOL
current. The presence of the SOL current does not modify the growth rate
(compare GS-SOLpf and GS cases). The GS-SOLpf-lt case is a modification to the
SOLpf case with $T_e=1$ eV at the edge of the SOL. The low edge temperature 
enhances the vacuum response \cite{Burke10,Ferraro10} and modestly modifies the
growth rate.
}
\label{gamma145098}
\end{center}
\end{figure}

In order to assess the impact, if any, on linear stability we examine the
toroidal-mode-number ($n_\phi$) growth-rate spectrum for shot 145098 at 1800 ms
with and without SOL-profile gradients. The reconstruction from shot 160414 is
during a stable inter-ELM period and thus it is not considered. As seen in
Fig.~\ref{gamma145098}, the growth rates are only minimally impacted by the 
inclusion of SOL-profile gradients where the
difference between the growth rates is at most 5\% (at $n_\phi=6$).  A
$72\times512$ mesh with bi-quintic elements is used for these computations.  The
growth rates with the SOL current are larger than those without for
$n_\phi<30$ and 1\% smaller at $n_\phi=30$. This is consistent with the effect
of enhanced resistivity and decreased density outside the LCFS as associated with
the SOL-profile gradients that leads the dynamics in this region to produce a
more vacuum-like response (see Refs.~\cite{Burke10,Ferraro10}) whereby
the low-$n_\phi$ modes are destabilized and the high-$n_\phi$ modes are
stabilized.  In order to
investigate this response further, we examine a case with a low electron
temperature, 1 eV, at the edge of the SOL region (annotated as GS-SOLpf-lt in
the figure). The stabilizing effect at low $n_\phi$ and destabilization at
high $n_\phi$ from the enhanced vacuum-like response is more apparent for this case.
Relative to the case without SOL current, the growth rates for this 
case are 7\% larger at $n_\phi=6$ and 8\% smaller at $n_\phi=30$. Thus the effect
of the SOL-profile gradients is modest compared to the effect of drift-stabilization 
(see e.g. Refs.~\cite{King16m,King14}) - a result that is consistent with prior 
PBM calculations~\cite{Snyder02}.

\begin{figure}
\begin{center}
\includegraphics[width=0.28\textwidth]{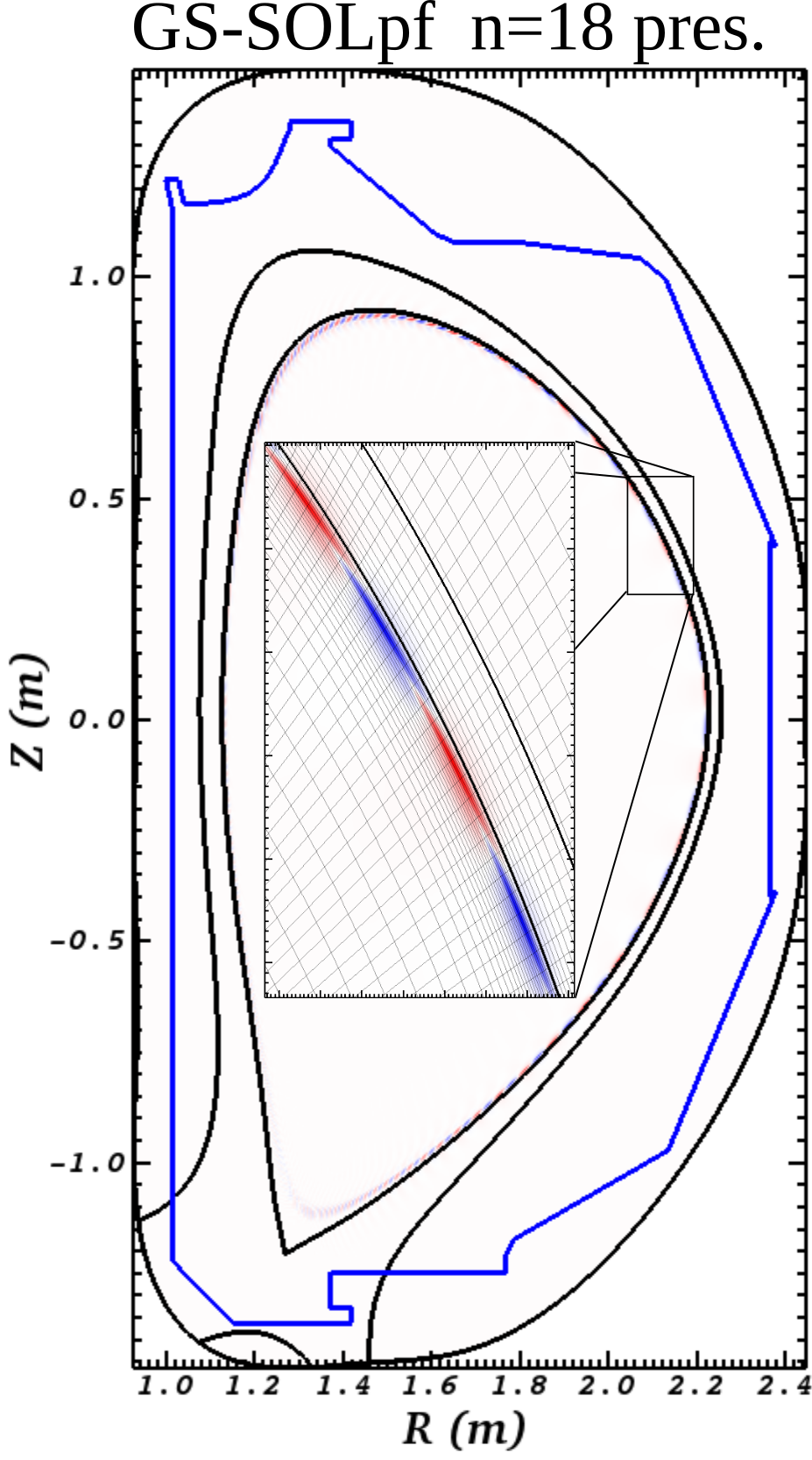}
\caption{ [color online]
Pressure contours (arbitrary units) from an extended-MHD calculation of a
$n_{\phi}=18$ mode during shot 145098 at 1800 ms where SOL
current is included in the equilibrium. The localization within the LCFS
is consistent with the small effect of the SOL current on the mode.
}
\label{pr145098}
\end{center}
\end{figure}

As seen in Fig.~\ref{pr145098}, the mode structure is localized within the
region just inside the LCFS. Convergence is affected by discontinuity in the
current-density profile when SOL-profile gradients are not included 
even though the mode is localized away from this discontinuity.
\textsc{nimrod} does not enforce the $\nabla \cdot B=0$ constraint through the
spatial discretization, but rather converges to a solution with small $\nabla
\cdot B$ error.  The $\ell^2$-norm of $\nabla \cdot B$ is reduced by
approximately 50\% in the cases with SOL-profile gradients relative to cases
without as a result of the continuous equilibrium profiles.
\section{Discussion and Conclusions} 
\label{sec:conclusions}

Relative to the effect on linear computations (Sec.~\ref{sec:pbm}),
the inclusion of SOL-profile gradients and associated continuous
current-density profiles has a greater impact on nonlinear modeling with
perturbations that are advected across the separatrix such as the studies of
QH-mode evolution of Ref.~\cite{King16qh}.  Nonlinear computations can be
affected by the inclusion of SOL-profile gradients in multiple ways: the
spatial resolution required to converge on the dynamics at the LCFS is less
with a continuous current profile as the dynamics are affected by
discontinuities in the current and by the changes in the vacuum-like response
as described in Sec.~\ref{sec:pbm}.

Perhaps more importantly, the methods described to extrapolate the
thermodynamic profiles in the SOL can be applied to modeling with flows.
Typically, the measured flows do not vanish at the LCFS and can be extrapolated
to zero in the SOL region.  Including this extrapolation both affects the
dynamics of the perturbations as they cross the LCFS and prevents the
computationally pathological case where perturbations may be advected quickly
inside the LCFS and not at all outside. Again, an example that applies these
methods to modeling with flows is found in Ref.~\cite{King16qh}.

The inclusion of SOL-profile gradients may have an effect outside the context
of modeling with initial value codes. The last-closed flux-surface locations
are shifted by up to $3\;cm$ in the cases included in this study.  While this
shift may not greatly affect MHD stability, it could have an impact on methods
that are predicated on highly accurate reconstructions such as RF injection for
current drive and/or tearing mode stabilization. These considerations may
motivate the inclusion of the SOL-profile gradients within the reconstruction
itself.

One limitation to the methods described here is that the flux from the plasma
that penetrates through the walls remains fixed. The flux can be decomposed
into plasma and external-coil contributions ($\psi = \psi_{plasma} + \psi_{ext.
coil}$).  A potential extension to this work is to perform a free-boundary
computation where $\psi_{ext. coil}$ is fixed but $\psi_{plasma}$ is allowed to
vary.  Additionally, $\psi_{ext. coil}$ could be extended to include
contributions from return currents flowing through the wall.  Ultimately, as
these computations become more sophisticated it may be better to include the
SOL-profile gradients in the $\chi^2$ minimization performed during the
reconstruction.

Even with this caveat, our methods represent substantial progress on the
initial condition for edge modeling. In particular, we have developed a
workflow whereby SOL-profile gradients and current can be included in the
initial condition for \textsc{nimrod} even if they are not included in the
reconstruction.  Using both global (e.g. total current) and local (e.g. the
separatrix location) metrics as well as a $\chi^2$ test we quantify the impact
of our methods on the accuracy of the initial condition after inclusion of
SOL-profile gradients.  We find that this impact is small and that the modified
initial condition closely resembles the state found by the reconstruction.
While linear stability is modestly impacted by the inclusion of the SOL-profile
gradients through an enhancement of the vacuum-like response, we argue that our
methods are more important for nonlinear modeling of dynamics across the
separatrix.
\appendix

\begin{acknowledgments} 
The authors would like to thank Alessandro Bortolon for generation of the EFIT
reconstruction of shot 160414 at 3025 ms, Jim Myra and Dan D'Ippolito
(deceased) for discussions of scrape-off-layer physics and ideas on how to
extend the profiles, Carl Sovinec and Eric Howell for the development of
\textsc{nimeq} and feedback on this development, and Stuart Hudson for use of
the \textsc{oculus} code and assistance on integration of the code into the
\textsc{nimrod} code base.  This material is based on work supported by US
Department of Energy, Office of Science, Office of Fusion Energy Sciences under
award numbers DE-FC02-06ER54875$^1$, DE-FC02-08ER54972$^1$ (Tech-X
collaborators), DE-FC02-04ER54698$^2$ (General Atomics collaborators) and
DE-FG02-03ER54692$^3$ (Auburn collaborators).
This research used resources of the National Energy Research Scientific
Computing Center, a DOE Office of Science User Facility supported by the Office
of Science of the U.S.  Department of Energy under contract
No.~DE-AC02-05CH11231.
\end{acknowledgments}
\bibliographystyle{apsrev4-1}
\bibliography{Biblio}

\end{document}